\documentclass[twocolumn,showpacs,preprintnumbers,amsmath,amssymb]{revtex4}

\usepackage{graphicx}
\usepackage{dcolumn}
\usepackage{bm}

\begin{document}

\title{Magnetic structure and phase diagram in a spin-chain system: Ca$_3$Co$_2$O$_6$}

\author{Yuri B. Kudasov}
\email{kudasov@ntc.vniief.ru}
\affiliation{Russian Federal Nuclear Center - VNIIEF, Sarov, 607188, Russia }

\date{\today}
\begin{abstract}
The low-temperature structure of the frustrated spin-chain compound Ca$_3$Co$_2$O$_6$
is determined by the ground state of the 2D Ising model on the triangular lattice.
At high-temperatures it transforms to the honeycomb magnetic structure. 
It is shown that the crossover between the two magnetic structures at 12~K arises from the entropy accumulated
in the disordered chains. This interpretation is in an agreement with the experimental
data. General rules for for the phase diagram of frustrated Ising chain compounds are formulated.
\end{abstract}


\maketitle

A combination of a low dimensionality 
and frustration in magnetic compounds, which contain weakly bound Ising spin-chains packed 
into a two-dimensional (2D) frustrated lattice, gives rise to a complex magnetic behavior. 
Antiferromagnetic (AFM) Ising chains form the triangular lattice 
in well-known Ising-chain compounds CsCoCl$_3$, CsCoBr$_3$ \cite{mekata1,mekata2,MI,shiba}. 
In these substances a partial AFM interchain order (honeycomb
magnetic structure) appears just below the ordering temperature. While the temperature decreases, 
two magnetic phase transitions occur to low-temperature ferrimagnetic phases. A theoretical explanation
of the magnetic structure of CsCoCl$_3$, CsCoBr$_3$ was proposed by Mekata in the framework of a mean-field theory 
\cite{mekata1}.
To describe the low-temperature phases an interaction between next nearest 
neighboring chains was included in the model.

Recently, Ca$_3$Co$_2$O$_6$ has drawn considerable attention to a step-like magnetization curve \cite{hardy1,
drillon, hardy2, hardy3, maignan}. The number of the steps in the curve depends strongly on a sweep rate of the
external magnetic field and temperature \cite{drillon, hardy2, maignan}. Two steps become apparent in the 
temperature
range from 12 K to 24 K \cite{maignan}. The first step takes place at the zero magnetic field. Then the magnetic
moment remains constant at about $1/3$ of the full magnetization up to the magnetic field of 3.6 T where the second
step occurs to the fully magnetized FM state.  At least four equidistant steps are clearly visible below 12 K at a
very low sweep rate. They are accompanied by a sizeable hysteresis. Similar phenomena were observed in other
spin-chain compounds, e.g. Ca$_3$CoRhO$_6$ \cite{nijtaka1, nijtaka2}.

The structure of Ca$_3$Co$_2$O$_6$ consists of Co$_2$O$_6$ chains running along the $c$ axis. The Ca ions are 
situated
between them. The chains are made up of alternating, face-sharing CoO$_6$ trigonal prisms and CoO$_6$ octahedra. 
The
crystalline electric field splits the energy level of Co$^{3+}$ ions into the high-spin ($S=2$) and low-spin 
($S=0$)
states.
The Co$^{3+}$ ions situated in the trigonal environment (CoI) are in the high-spin state and the octahedral Co 
sites
(CoII) occurs in the low-spin state. In the last case the energy difference between the low-spin and high-spin 
states
is very small and a tiny fraction of CoII sites is reported to be in the high-spin state. The crystalline electric
field leads also to a very strong Ising-like anisotropy at the CoI sites. The chains form triangular lattice in the
$ab$ plane that is perpendicular to the chains. An in-chain exchange interaction between magnetic CoI ions through 
the
octahedra with non-magnetic CoII ions is ferromagnetic (FM). The parameter of the FM in-chain coupling ($J_F$) was
found from the magnetic susceptibility at high temperatures \cite{kageyama}, specific heat \cite{hardy3}, and 
theoretical
calculations \cite{fresard}. These estimations are in a reasonable agreement with each other ($J_F \approx 25$K). 
The interchain interaction is
antiferromagnetic (AFM) and much weaker than the in-chain one. 
It should be mentioned that the topology of the magnetic sublattice in Ca$_3$Co$_2$O$_6$ is much more complex
than that in CsCoCl$_3$, e.g. there exist helical paths \cite{fresard}.
A partial AFM order of chains appears at $T_{C1}$=24~K.
A weak feature concerned most probably with a transition in a new interchain order was also observed at around
$T_{C2} \approx 12$~K. This scenario of magnetic interactions in Ca$_3$Co$_2$O$_6$ is consistent with results of 
x-ray
photoemission spectroscopy \cite{x-ray}, neutron scattering \cite{neutron, petrenko}, magnetization  and specific 
heat
measurements \cite{hardy1, drillon, hardy2, hardy3, maignan,martinez}, nuclear magnetic resonance \cite{nmr}, and
theoretical calculations of indirect interactions between CoI sites \cite{fresard}.

In the previous paper \cite{kudasov} I proposed a model to interpret the step-like magnetization curve of 
Ca$_3$Co$_2$O$_6$ at very low temperatures. It was based on two assumptions. (i) The 
full in-chain FM order was supposed (rigid chains), i.e. only two state of a 
chain (spin-up and spin-down) were considered. (ii) Since Ca$_3$Co$_2$O$_6$ and related compounds demonstrate
very slow dynamics of chain re-orientation, the system was assumed to evolve over metastable states with a 
variation of the external magnetic field.
A single-flip technique \cite{kim} was used to determine the metastable states as local minima.
The ground state of 2D Ising model on the triangular lattice \cite{wannier} was taken as the
initial state at the zero magnetic field. This approach
allowed explaining the main features of the low-temperature step-like magnetization 
curve. At the same time, it is obvious that the first assumption is valid for the zero temperature only. In the 
present
paper the rigid-chain model \cite{kudasov} is extended to low non-zero temperatures to clarify the origin of the 
transition from
the low- to high-temperature magnetic structure at 12 K. 

\begin{figure*}
\includegraphics{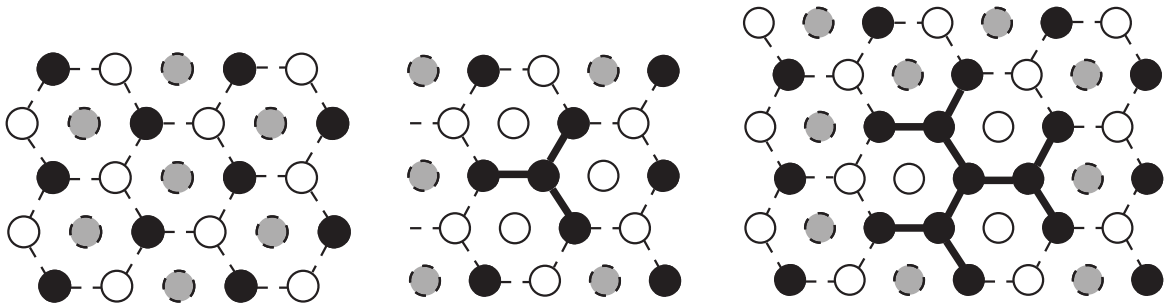}
\caption{\label{f1} The magnetic structures of the triangular lattice of chains: (left) the honeycomb structure,
(middle) the
isolated tripod configuration, and (right) the three tripod connected in the star arrangement. The black and white
circles are spin-up and spin-down states, correspondingly. The gray circles can be either spin-up or spin-down
states in the low-temperature phase or disordered chains in the high-temperature phase.}
\end{figure*}

Let us start with the rigid-chain model \cite{kudasov} at the zero temperature. 
The ground state is described by the Ising model 
on the triangular lattice and can be represented through a 
sequence of approximations \cite{wannier,kudasov}. The honeycomb structure shown in the left panel of Fig.\ref{f1} 
is the first one. 
It can be improved by including an isolated tripod configuration (the middle panel of Fig.~\ref{f1}), conjugate 
structures of two tripods, 
three tripods in a star configuration (the right panel of Fig.~\ref{f1}), and etc. In all the structures there 
exist chains with the zero
effective field, that is, with three spin-up and three spin-down nearest neighbors. When the temperature is 
non-zero
these chains should be in a disordered state. This fact breaks the first assumption of the rigid chain model.
In a real crystal the chains are of a large but finite length. The upper limit of the 
number of atoms in a chain hardly exceeds 
$L=10^3$ because it is difficult to obtain the purity of a complex compound better than 
99.9~$\%$. In addition, taking into
account structural defects we obtain $L=10^2$ as a reasonable estimation of 
the average chain length.
Fortunately, it is shown below that the final results depend very weakly on the chain length. 

The partition function
of the Ising chain at the zero effective field has a well-known form (e.g. \cite{stanley})
\begin{equation}
Z_c=2^L \cosh^{L-1}K
\label{Z_chain}
\end{equation}
where $K=J_F/T$, $T$ is the temperature, and $J_F>0$ for the FM in-chain coupling. 
A two-spin correlation function determined from Eq.(\ref{Z_chain}) is $\Gamma(r)=\tanh^{r}K$ where $r$
is the distance. Therefore, if $\Gamma(L)\approx 1$ there is a large probability of the 
full FM in-chain order. This condition can be written as 
\begin{equation}
K>>\frac{1}{2}\ln(L). 
\label{cross_1}
\end{equation}
It gives the crossover temperature at about 11~K 
that is close to the $T_{C2}$. Above this
temperature $\Gamma(L) << 1$ and we can't consider the chains at the zero effective field as rigid ones
because they are in the disordered state.

The most part of chains is under non-zero effective field. For $i$-th chain it can be written as
\begin{equation}
h_i = J_A \sum_{j(i)}{<\sigma_j^z>}-B
\label{h_eff}
\end{equation}
where $j(i)$ denotes summation over the nearest-neighbors of the $i$-th chain, $J_A<0$
is the parameter of the AFM interchain coupling, $<\sigma_j^z>$ is the average magnetic 
moment of the $j$-th chain, and $B$ is the external magnetic field. 
Under the effective field the partition function 
can be found exactly for the Ising ring (a chain with periodic boundary conditions) by 
means of a transfer matrix technique \cite{stanley}
\begin{equation}
Z_c(H)=\lambda_{+}^{L}+\lambda_{-}^{L}
\label{Z_chain_h}
\end{equation}
where 
\begin{equation*}
\lambda_{\pm}=e^{K}\cosh(H) \pm \sqrt{e^{2K}\sinh^2(H)+e^{-2K}}
\end{equation*}
and $H_i=h_i/T$. This equation gives the partition function of the $i$-th single chain. 
For the sake of simplicity we have omitted the low indexes here. 
Due to the periodic
boundary conditions Eq.~(\ref{Z_chain_h}) 
gives a different limit at $H \rightarrow 0$ as compared to that of Eq.~(\ref{Z_chain}). However, both 
the expressions converge with increase of $L$. For very weak effective fields, i.e.
$\sinh(H)\sim{H}<<\exp(-2K)$, we obtain from Eq.~(\ref{Z_chain_h}) the same condition for
the FM in-chain order given by Eq.~(\ref{cross_1}). If the effective field is not extremely small, i.e.
$\sinh{H}>>\exp(-2K)$, the full FM order appears
when $\tanh(H)\sim{2H}>>L^{-1}$. We see that in any case a small effective field is sufficient for
the crossover to the FM in-chain order.

The foregoing discussion shows that at low temperatures we can assume the full FM
in-chain order in all the chains with the exception of the chains with the zero 
or very small effective fields. Returning to Fig.\ref{f1} the state of the grey chains 
is rigid or disordered depending on the temperature. 
In fact this assumption is valid in a wide temperature range and broken only in a narrow region close to $T_{C1}$. 
Then the partition function of the honeycomb structure at the zero external magnetic 
field can be represented as
\begin{equation}
Z_{hc}=Z_c^{N/3}e^{-E_0/T} 
\label{Zhc}
\end{equation}
where $N$ is the number of chains and $E_0$ is the ground state energy of the honeycomb structure. At the zero 
temperature this expression
gives the entropy $S_{hc}=(N/3)\ln 2\approx 0.231 N$. The tripod configuration appears
at this limit with the probability $1/12$ per chain and the entropy increases
up to $S_t=(5N/12) \ln 2\approx 0.289 N$ \cite{wannier}. At non-zero temperatures 
the tripod appears if the three chains in the centers of adjacent hexagons are in an ordered state,
the spin-down state (white circles) in the case shown in the middle panel of Fig.~(\ref{f1}).
Thus, the probability of tripod configuration per chain can be written as 
\begin{equation}
P_t=\left[\frac{e^{K(L-1)}}{Z_c}\right]^3=\left[2(1+e^{-2K})^{L-1}\right]^{-3} 
\label{tripod}
\end{equation}
Taking into account the the tripod configurations the partition function of the triangular 
lattice of the chains is given by
\begin{equation}
Z=Z_c^{N[1/3+P_t]}e^{-E_0/T} 
\label{Z}
\end{equation}
It should be mentioned that the tripods do not exhaust all the configurations in the 
ground state of the 2D Ising model on the triangular lattice but they give a 
good approximation as one can see comparing $S_t$ with the exact value of the
entropy of the triangular lattice $S=0.3231 N$ \cite{wannier}.

From Eq.~(\ref{Z}) one can easily calculate the free energy and thermodynamic properties
of the system.
At the zero temperature this equation leads to the entropy $(5N/12)\ln 2$. While the temperature
grows $P_t$ decreases and, when $[1+\exp(-2K)]^{L-1}>>1$, the second term in the square brackets 
of Eq.~(\ref{Z}) becomes negligible.
Then Eq.~(\ref{Z}) coincides with the partition function of the honeycomb 
structure (\ref{Zhc}). 
The crossover from the low- to high temperature regime occurs when $\exp(-2K)(L-1) \approx 1$. 
This gives an estimation of the crossover temperature $T_{C2}\approx$11~K very close to the experimental value.
We also see that the dependence of the crossover temperature on the chain length is logarithmic and, therefore, 
rather weak.
 
Now we have a complete picture of the phase diagram in Ca$_3$Co$_2$O$_6$. At very low temperatures, i.e. much 
below $T_{C2}$, all the chains are rigid, that is, with the full FM in-chain order. The interchain magnetic
order is the same as in the Ising model on the triangular lattice. While temperature increases the in-chain FM
order collapses in the chains that are under the zero effective field. Whereas the entropy at the zero temperature
is due to various configurations on the 2D triangular lattice, at higher temperature the entropy is
accumulated mainly in disordered chains. That is why, the statistical weight of configurations with a larger number 
of
disordered chains quickly grows with temperature. Since, as one can see from Fig.~\ref{f1}, the honeycomb structure 
have the 
largest number of disordered chains, it dominates above $T_{C2}$ and the tripod configurations are suppressed. 
From this consideration we also see that the crossover at $T_{C2}$ is not a real phase transition. That is why,
there is no peak or other features observed in the temperature dependence of the specific heat at the crossover
temperature \cite{hardy3}. On the other hand, the crossover is rather sharp because the statistical weight of the 
tripod
configurations depends on the temperature exponentially. 

\begin{figure}
\includegraphics{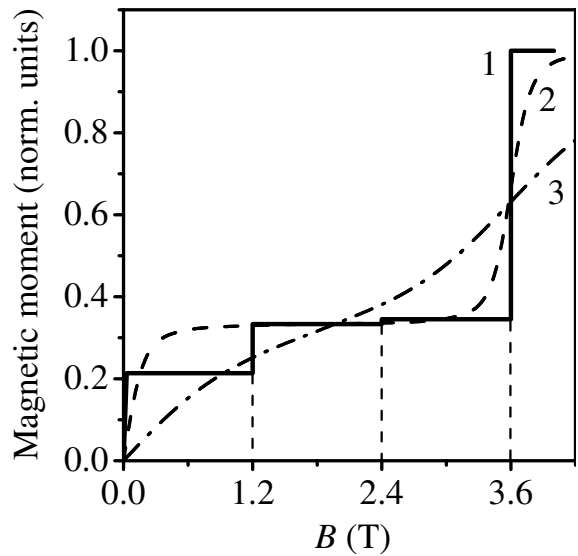}
\caption{\label{f2} The magnetic moment as a function of the magnetic field. The solid line is the magnetization
curve at a very low temperature \cite{kudasov}, the dash line 2 and  dash-dot line 3 are calculated for the 
honeycomb
structure at $T=$12.5~K and $T=$20~K correspondingly.}
\end{figure}

Our interpretation of the magnetic structure transformation agrees with the experimental observations.
The two steps that are observed in the magnetization 
curve above $T_{C2}$ \cite{maignan} together with the neutron scattering data \cite{petrenko} indicate clearly the 
honeycomb magnetic structure. Using Eq.~(\ref{Z_chain_h}) we calculated the magnetization curves
at high temperature for the honeycomb structure as a initial state (curves 2 and 3 in Fig.~\ref{f2}). 
They fit the experimental curves \cite{maignan}. This means that
the smoothing of the steps at high temperatures is due to the magnetization of the
disordered chains. On the contrary, the four steps calculated at the low temperature
(Fig.~\ref{f2}) are very sharp and some smoothing observed experimentally \cite{hardy2} stems
most probably from the slow dynamics of the chains re-orientation.

The magnetic phase diagram determined in the present work at $H=0$ is quite different from that of Ref.~\cite{MI}.
This discrepancy appears because of two reasons mainly. (i) As was mentioned above the structure of the magnetic 
sublattice in Ca$_3$Co$_2$O$_6$ is more complex as compared to CsCoCl$_3$. (ii) In Ref.~\cite{MI} the in-chain 
exchange parameters was assumed to be equal to or even less then the interchain one. 
This gives rise to the randomly modulated phase (RMP) in a wide temperature region. Fluctuations in the RMP develop 
along both $ab$-plane and $c$-axis. 
In Ca$_3$Co$_2$O$_6$ the in-chain exchange parameter is much stronger than the interchain one. The RMP
is suppressed in this case and the rigid honeycomb structure is stable up to high temperatures. 
Well above the crossover temperature our approach coincides with Shiba's mean-field
model \cite{shiba}. However, it can't be applied to the crossover region because all the phase boundaries in this
model are phase transitions in contrast to the results obtained above and experimental data.

The consequences of the results obtained are quite general. 
We can formulate universal rules for the phase diagram of frustrated Ising chain
compounds. (i) The low-temperature magnetic phase is formed by rigid chains and
its ground state is determined by the ground state of the 2D Ising model.
It should be mentioned that in case of Mekata's theory of CsCoCl$_3$ the
ground state of the 2D Ising model is different from that considered in the 
present paper because of the additional interaction between next-nearest-neighbor chains.
(ii) In the high-temperature magnetic structure the number of disordered chains at the zero
effective field that should be maximal to increase the entropy and lower the free energy. 
That is why, at high temperatures the magnetic structure of
the triangular lattice of Ising chains tends to the honeycomb one. 
These rules are valid for the FM and AFM Ising chain 
systems Ca$_3$Co$_2$O$_6$, CsCoCl$_3$, CsCoBr$_3$ as well Ising chain compounds
with distortions, e.g. TlCoCl$3$ \cite{TlCoCl3} and RbCoBr$3$ \cite{RbCoBr3}.

I am grateful to V. Uzdin for fruitful discussions.
This study was partially supported by INTAS grant (03-51-4778) "Hierarchy of scales 
in magnetic nanostructures".

\bibliography{Yura}

\end{document}